\documentclass[twocolumn]{jpsj2}

\title{
Excitonic versus electron-hole liquid phases in Tm[Se,Te] compounds
}

\author{
Franz X. Bronold$^1$\thanks{E-mail address: bronold@physik.uni-greifswald.de},
Holger Fehske$^1$, and Gerd R\"opke$^2$}

\inst{
$^1$Institut f\"ur Physik, Ernst-Moritz-Arndt-Universit\"at Greifswald, 
D-17489 Greifswald, Germany\\
$^2$Fachbereich Physik, Universit\"at Rostock, 
D-18051 Rostock, Germany
}

\recdate{\today}

\abst{
We discuss, from a theoretical point of view, excitonic phases at 
the pressure induced semiconductor-semimetal transition in $\rm TmSe_{0.45}Te_{0.55}$,
focusing, in particular, on the stability against an electron-hole liquid. The 
electron-hole pair density parameter $r_s(E_g,T)$ is calculated within the quasi-static 
plasmon pole approximation as a function of temperature $T$ and energy gap
$E_g$ and converted into $-E_g(r_s,T)$. A comparison of this quantity, which is the
electron-hole pair chemical potential, with 
the exciton binding energy reveals that excitons should be suppressed in contrast 
to experimental evidence for excitonic phases. We suspect therefore inter-valley 
exciton scattering and exciton-phonon scattering to substantially stabilise excitons
in $\rm TmSe_{0.45}Te_{0.55}$. 
}

\kword{%
semiconductor-semimetal transition, electron-hole liquids, exciton condensation
}

\begin{document}
\maketitle

\section{Introduction}
The possibility of an excitonic insulator -- understood as a 
macroscopic, phase coherent quantum state (exciton condensate) -- separating 
below a critical temperature a semiconductor (SC) from a semimetal (SM), has been 
anticipated by theorists long time ago~\cite{HR68}. Experimental efforts, however,
to establish this phase in real compounds largely failed. The most 
promising candidates so far are the hexaborides~\cite{YHT99}, where the unexpected 
ferromagnetism may be interpreted in terms of a doped excitonic 
insulator~\cite{ZRA99,BBM02}, 
and ${\rm TmSe_{0.45}Te_{0.55}}$, where the unusual thermal diffusivity may be 
due to a superfluid exciton state~\cite{WBM04}. 

We focus on ${\rm TmSe_{0.45}Te_{0.55}}$.
In a series of experiments, Wachter and coworkers compiled strong evidence 
for excitonic phases in the vicinity of the pressure-induced SC-SM  
transition in ${\rm TmSe_{0.45}Te_{0.55}}$~\cite{NW90,W01,WBM04}. In particular, 
the anomalous increase of the electrical resistivity in a narrow pressure range 
around 8~kbar indicated the appearance of a new phase below 250~K.
Wachter and coworkers suggested that this new phase might be an ``excitonic insulator'',
and, assuming the energy gap $E_g$ to decrease with increasing pressure, constructed a 
phase diagram for ${\rm TmSe_{0.45}Te_{0.55}}$ in the ${E_g-T}$ plane~\cite{NW90}. Later
they found in the same material a linear increase of the thermal diffusivity below
20~K and related this to a superfluid exciton state~\cite{WBM04}. Both excitonic
phases are located on the SC side of the SC-SM transition in ${\rm TmSe_{0.45}Te_{0.55}}$.

In our previous work~\cite{our} we presented a theoretical analysis of 
these astonishing experimental findings. We calculated for a model mimicking 
the situation in ${\rm TmSe_{0.45}Te_{0.55}}$ the phase boundary $T_c(E_g)$
for an excitonic insulator and investigated the composition of its environment. 
Our main conclusions were: (i) The phase boundary constructed from the resistivity 
data does not embrace the excitonic insulator. Instead, it most probably traces the 
exciton-rich region above an excitonic insulator, where excitons dominate the total 
density, provide abundant scattering centres for the charge carriers, and induce the 
observed resistivity anomaly. (ii) The linear increase of the thermal diffusivity 
below 20~K, on the other hand, could signal exciton condensation. The large
electron-hole mass asymmetry places the excitonic insulator entirely on the SC side, 
where it is supported by strongly bound, bosonic excitons. A transition to a 
superfluid state, similar to the transition in liquid He~4, is thus conceivable and 
could give rise to the observed atypical thermal diffusion.

We based our calculation on an effective-mass, Wannier-type model for electrons 
in the lowest conduction band (CB) and holes in the highest valence band (VB),
with an indirect energy gap separating the two bands, a multiplicity factor to 
account for three equivalent CB minima arising from the fcc crystal structure 
of ${\rm TmSe_{0.45}Te_{0.55}}$, and a statically screened Coulomb
potential operating between the constituents of the model (in contrast to 
the quasi-static screening model used below). Because
of the limitations of the model, which, for instance, neglects 
lattice effects, we could not obtain complete quantitative agreement between
theory and experiment. Yet, our results indicate that 
Wachter and coworkers~\cite{NW90,W01,WBM04} may have seen an
exciton condensate. 

Clearly, within an effective electron-hole model quantitative agreement between 
experiment and theory cannot be expected. Instead of augmenting the model with  
additional degrees of freedom, we still want to stay in the present paper within 
its bounds and discuss the possibility of an electron-hole liquid~\cite{R77} which 
competes with excitonic phases in materials where the band structure is 
multi-valleyed. The results will further benchmark the model and thus guide 
its modification.

\section{Method}
The thermodynamics of an electron-hole liquid is contained in the electron-hole pair 
chemical potential $\mu_{eh}(r_s,T)$ which depends on the electron-hole pair density 
parameter $r_s$ and the temperature $T$~\cite{R77}. In our case, 
$\mu_{eh}=-E_g$, with $E_g$ the energy gap fixed by the external pressure. Therefore,
we first have to determine $r_s(E_g,T)$ and then convert this relation into 
$-E_g(r_s,T)$. Comparing $-E_g(r_s,T)$ with the exciton binding energy and
analysing for fixed temperature its $r_s$-dependence enables us to determine
the region in the $E_g-T$ plane where an electron-hole liquid may be favoured over 
excitonic phases. We calculate $-E_g(r_s,T)$ in the quasi-static plasmon-pole 
approximation~\cite{HS84}.

\begin{figure}[t]
\begin{center}
\includegraphics[width=0.9\linewidth]{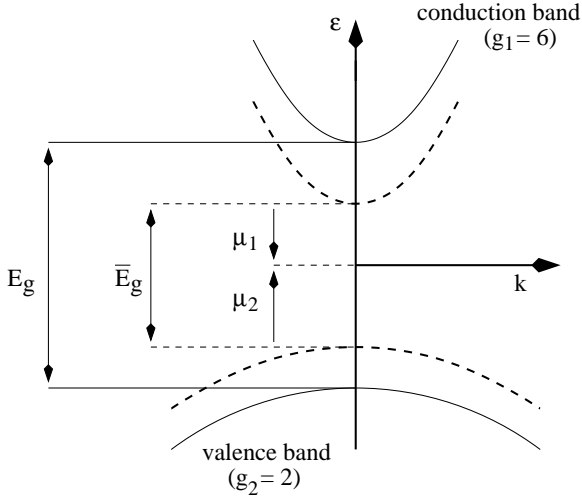}
\end{center}
\caption{The band structure of the effective-mass, Wannier-type model. We 
measure momenta ${\bf k}$ for the CB electrons and VB holes from the respective 
extrema of the bands which, in reality, are separated by ${\bf w}={\bf K}/2$
where ${\bf K}$ is the reciprocal lattice vector connecting the $\Gamma$ with 
the $X$ points of the  Brillouin zone for a fcc lattice. The energy 
gap $E_g$ varies continuously through zero under pressure. Solid 
(dashed) parabola denote the bare (renormalised) energy bands.
Note, the auxiliary chemical potentials $\mu_i$ are defined with respect
to the shifted band edges. The factors $g_i$ account for spin
and the number of equivalent valleys. 
}
\label{f1}
\end{figure}

The schematic band structure used in Ref.~\cite{our} to model 
${\rm TmSe_{0.45}Te_{0.55}}$
is shown in Fig.~\ref{f1}. Within the quasi-static plasmon-pole approximation
the CB electron ($i=1$) and VB hole ($i=2$) selfenergies, leading, at finite
filling, to the renormalization of the energy bands 
(dashed parabola in Fig.~\ref{f1}), 
can be written as 
\begin{equation}
\Sigma_{i}({\bf k})=\Sigma^{sx}_{i}({\bf k})+
\Sigma^{ch}_{i}({\bf k})~,
\end{equation}
where the first term, 
\begin{equation}
\Sigma^{sx}_{i}({\bf k})=-\sum_{\bf k'}V_s({\bf k}-{\bf k'})
n_{\rm F}(\epsilon_i({\bf k'})-\mu_i)~, \label{SX}
\end{equation}
denotes the exchange energy and the second, 
\begin{equation}
\Sigma^{ch}_{i}({\bf k})=\frac{1}{2}\sum_{\bf k'}
\big[V_s({\bf k'})-V_0({\bf k'})\big]~, \label{CH}
\end{equation}
the Coulomb-hole energy. Here, $V_0({\bf q})=4\pi e^2/\varepsilon_0 q^2$ is the  
Coulomb potential reduced by the background dielectric constant $\varepsilon_0$, 
$V_s({\bf q})$ is the statically screened Coulomb potential to be defined 
below, $\epsilon_i({\bf k})={\bf k}^2/2m_i$ ($m_i$ are the effective masses), 
and $n_{\rm F}(\epsilon)$ is the Fermi 
function. In contrast to the static approximation used in Ref.~\cite{our}, which 
replaces directly in the Hamiltonian $V_0({\bf q})$ by $V_s({\bf q})$,
the selfenergies contain now not only the exchange corrections but also the 
Coulomb-hole. 

The screened Coulomb potential in Eqs.~(\ref{SX}) and (\ref{CH}) is the 
static limit of the dynamically screened Coulomb potential, 
\begin{equation}
V_s({\bf q})\equiv V_s({\bf q},0)=V_0({\bf q})/\varepsilon({\bf q},0)~,
\end{equation}
where the dielectric function is given by 
\begin{equation}
\frac{1}{\varepsilon({\bf q},\omega)}=1+
\frac{\omega_{pl}^2}{(\omega+i\eta)^2-\omega({\bf q})^2}~,
\end{equation}
with an effective plasmon dispersion, 
\begin{eqnarray}
\omega({\bf q})^2=\omega_{pl}^2\big[1+\big(\frac{q}{q_s}\big)^2\big] + 
C\big(\frac{q^2}{4m_1}+\frac{q^2}{4m_2}\big)^2~,
\label{plasmon}
\end{eqnarray}
constructed in such a way that sum rules are satisfied~\cite{HS84}.
When the plasmon-pole approximation is constructed from the Lindhard dielectric function, 
the coefficient $C=1$. It can be however also used as a fit parameter to correct for
the overestimation of screening within the quasi-static approximation. In that case 
$1\le C \le 4$~\cite{HS84}. The screening wave number,
\begin{equation}
q_s=\sqrt{\frac{4\pi e^2}{\varepsilon_0}\big[\frac{\partial}{\partial \mu_1}n_1
+\frac{\partial}{\partial \mu_2}\bar{n}_2\big]}~,
\label{qs}
\end{equation}
and plasma frequency,
\begin{equation}
\omega_{pl}=\sqrt{\frac{4\pi e^2}{\varepsilon_0}\big[\frac{n_1}{m_1}+\frac{\bar{n}_2}{m_2}\big]}~,
\label{wpl}
\end{equation}
entering Eq.~(\ref{plasmon}) depend on the CB electron and VB hole densities given, 
respectively, by 
\begin{subequations}
\begin{align}
n_1 &= g_1 \sum_{\bf k} n_{\rm F}(\epsilon_1({\bf k})-\mu_1)~, \label{n1}\\
\bar{n}_2 &= g_2 \sum_{\bf k} n_{\rm F}(\epsilon_2({\bf k})-\mu_2)~,
\label{n2}
\end{align}
\end{subequations}
where the multiplicity factors $g_i$ contain the spin and the number 
of equivalent valleys. 

The main effect of the selfenergies $\Sigma_{i}$ is a rigid, ${\bf k}$-independent
shift of the single particle dispersions leading to a renormalised band gap
$\bar{E}_g$. This situation has been anticipated in Eqs.~(\ref{SX}), (\ref{n1}),
and (\ref{n2}), where we introduced auxiliary chemical potentials $\mu_i$,  
measured from the shifted band edges (see Fig.~\ref{f1}). By 
construction, 
\begin{equation}
\mu_1+\mu_2=-\bar{E}_g~.
\label{Egbar}
\end{equation}
The bare energy gap $E_g$, that is, the parameter which is experimentally 
controlled via pressure, is given by
\begin{equation}
E_g=\bar{E}_g-\Sigma_1(0)-\Sigma_2(0)~.
\label{Eg}
\end{equation}
Supplementing  Eqs.~(\ref{Egbar}) and (\ref{Eg}) with the condition of 
charge neutrality, 
\begin{equation}
n_1=\bar{n}_2~,
\label{neutrality}
\end{equation}
we obtain a closed set of equations for the three unknown parameters
$\mu_1$ and $\mu_2$, and $\bar{E}_g$. 

To construct from Eqs.~(\ref{Egbar})~--~(\ref{neutrality}) the 
electron-hole pair chemical potential $-E_g(r_s,T)$,
we proceed as follows: First, we calculate from Eqs.~(\ref{neutrality}), (\ref{Egbar}), 
(\ref{n2}) and (\ref{n1}), the auxiliary chemical potentials $\mu_i$ as a
function of $\bar{E}_g$ and $T$. Inserting $\mu_i(\bar{E}_g,T)$ in Eqs.~(\ref{qs}) and 
(\ref{wpl}), we obtain $q_s(\bar{E}_g,T)$ and $\omega_{pl}(\bar{E}_g,T)$, which enables 
us to determine $V_s$ and thus the selfenergies $\Sigma_i$ in the whole $\bar{E}_g-T$ 
plane. Via Eq.~(\ref{Eg}) we then obtain the important relation $E_g(\bar{E}_g,T)$. Finally, 
we use either Eq.~(\ref{n1}) or Eq.~(\ref{n2}) to map, for a given temperature $T$, the 
renormalised energy gap $\bar{E}_g$ on to the electron-hole pair density 
$n\equiv n_1=\bar{n}_2$. 
Using the density parameter $r_s=(3/4\pi n \cdot a^3_X)^{1/3}$, with $a_X$ the exciton 
Bohr radius defined in the next section, instead of the pair density, we find $-E_g(r_s,T)$.

\section{Results}

In the following, we 
measure energies and lengths in units of the exciton Rydberg $R_X=1/2ma_X^2$ and the
exciton Bohr radius $a_X=\varepsilon_0/me^2$, respectively ($\hbar=1$). The model shown   
in Fig.~\ref{f1} is then completely specified by the multiplicity factors $g_i$ 
and the effective mass ratio $\alpha=m_1/m_2$. For ${\rm TmSe_{0.45}Te_{0.55}}$ the 
corresponding values are $g_1=6$, $g_2=2$, and $\alpha\approx 0.015$~\cite{WBM04,our}. 
Due to the lack of data about band gap renormalization in $\rm TmSe_{0.45}Te_{0.55}$, 
we cannot use $C$ in Eq.~(\ref{plasmon}) as a fit parameter. Thus,   
$C=1$, which almost certainly overestimates screening. 
\begin{figure}[t]
\begin{center}
\includegraphics[width=0.9\linewidth]{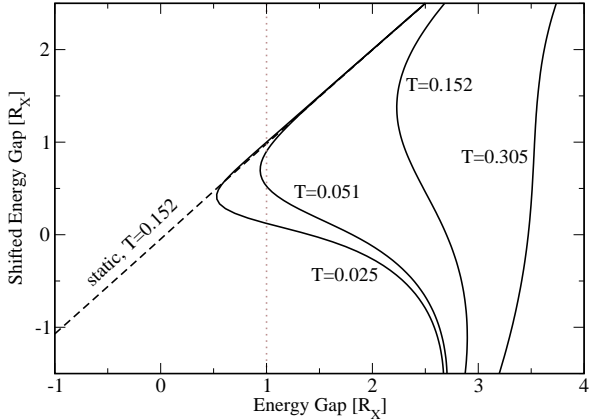}
\end{center}
\caption{Shifted energy gap $\bar{E}_g$ as a function of the
pressure-tuned energy gap $E_g$ and temperature $T$ for
$\rm TmSe_{0.45}Te_{0.55}$ parameters: $\alpha=0.015$, $g_1=6$,
and $g_2=2$. The dotted line denotes the energy where the
excitonic instability would occur at $T=0$ and the dashed line
gives $\bar{E}_g(E_g,T=0.152)$ in the static approximation.
}
\label{f3}
\end{figure}

The shifted energy gap $\bar{E}_g$ as a function of the energy gap $E_g$ is shown 
in Fig.~\ref{f3} for several temperatures. First we note the drastic difference 
to the static approximation ($V_0\rightarrow V_s$ in the Hamiltonian). Whereas 
in the static approximation $\bar{E}_g$ decreases monotonously with $E_g$ for 
all temperatures, the quasi-static approximation gives, for some $E_g$-range,
a triple-valued $\bar{E}_g$ for $T<0.305$. Because of the one-to-one 
correspondence between $\bar{E}_g$ and the pair density $n$, this means 
the system cannot choose a unique pair density. As a consequence,
there may be parts in the $\bar{E}_g-T$-plane where an electron-hole gas 
coexists with an electron-hole liquid. Note, the renormalised band gap
$\bar{E}_g$ resembles the selfenergy parameter $S$ introduced before 
to set up a Hartree-Fock description of the gas/liquid
transition in a generic electron-hole plasma~\cite{S77}.

\begin{figure}[t]
\begin{center}
\includegraphics[width=0.9\linewidth]{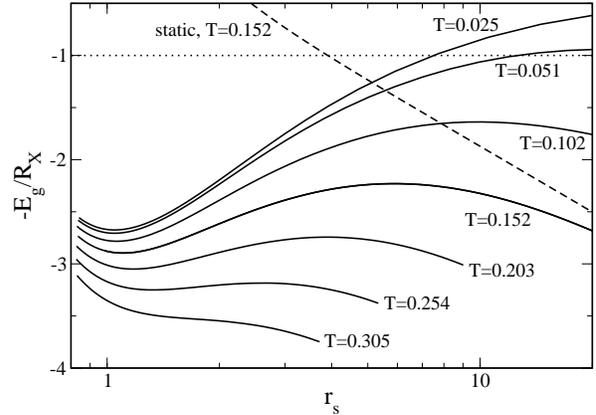}
\end{center}
\caption{Electron-hole pair chemical potential $-E_g(r_s,T)$ as a
function of electron-hole pair density parameter $r_s$ and
temperature $T$ for $\rm TmSe_{0.45}Te_{0.55}$ parameters:
$\alpha=0.015$, $g_1=6$, and $g_2=2$.
The dotted line indicates where the
excitonic instability would occur at $T=0$ and the dashed line
gives $-E_g(r_s,T=0.152)$ in the static approximation. Note, for
$T<0.305$, the isotherms go through a minimum.
}
\label{f4}
\end{figure}

The corresponding pair chemical potential $-E_g(r_s,T)$ is shown in Fig.~\ref{f4}.
Various subtle points should be noticed before we can attempt an interpretation of 
the data. First, $-E_g(r_s,T)$ does not asymptotically merge with the exciton
Rydberg for $r_s\rightarrow \infty$. Thus, the results in the extreme dilute limit are 
flawed, not only 
in the quasi-static approximation (solid lines) but also in the static approximation 
(dashed line). This is a well-known shortcoming of random-phase-approximation based 
screening models, which neglect multiple-scattering between charged carriers. 
Including multiple-scattering (ladder diagrams) in the quasi-static (static)
approximation would force the 
chemical potential $-E_g(r_s,T)$ to approach for $r_s\rightarrow \infty$ 
the exciton Rydberg from below
(above). Second, the densities of interest are rather
low ($r_s\gtrsim 1$). Corrections due to multiple scattering are therefore not  
necessarily small and should be included, a formidable task
beyond the scope of the present paper.

With these caveats in mind we now extract the physical implications of the data 
shown in Fig.~\ref{f4}. As expected from the $\bar{E}_g$-isotherms, the model
shown in Fig.~\ref{f1} yields for $\rm TmSe_{0.45}Te_{0.55}$ parameters
a chemical potential which shows the typical van-der-Waals behaviour: Above a 
critical temperature ($T_{crit}\approx 0.254$), the chemical potential increases 
monotonously with density (recall: $r_s\sim 1/n^{1/3}$). Below the critical   
temperature, the chemical potential has a local maximum and a local 
minimum. This can be most clearly seen from the $T=0.203$ isotherm. A 
Maxwell construction would thus give the equilibrium chemical potential and
the coexistence region between the gaseous and the liquid phases. Anticipating
however the correct dilute limit, a Maxwell construction cannot be made and the 
minimum simply indicates a jump in the pair densities, that is, for large enough 
energy gaps the pair density $n$ drops to zero~\cite{ZR00}. Only when 
the exciton binding energy is 
below the minimum of the chemical potential, a gas/liquid transition 
and with it a coexistence region could appear. In that case, excitons 
would be also stable in a certain range. Because of the experimental evidence 
for excitonic phases, we expect this situation to be actually realised in 
$\rm TmSe_{0.45}Te_{0.55}$.

The minimum of the pair chemical potential $-E_g(r_s,T)$ is below 
the exciton Rydberg. 
Therefore, taking the
correct $r_s\rightarrow\infty$ limit of $-E_g(r_s,T)$ into account, 
our results for the pair chemical potential suggest
that in the Wannier-type model excitons and thus excitonic phases 
(exciton gas, excitonic insulator) are 
unstable against a metallic phase. Anticipating, in addition, the decrease of the 
exciton binding energy with increasing density, the metallic phase would be in fact
favoured over most of the relevant parameter range discussed 
in Ref.~\cite{our} and in stark contrast to the experimentally observed increase
of the resistivity. 

\begin{figure}[t]
\begin{center}
\includegraphics[width=0.9\linewidth]{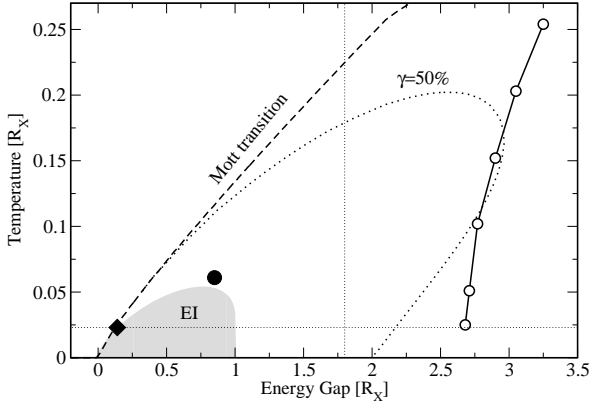}
\end{center}
\caption{Phase boundaries obtained for
$\rm TmSe_{0.45}Te_{0.55}$ parameters: $\alpha=0.015$, $g_1=6$,
and $g_2=2$. The thick dotted line is the $50\%$-contour 
of the bound state fraction $\gamma$~\cite{our}.
Beyond the dashed line excitons disappear because of the Mott
effect and the grey region denotes the excitonic
insulator (EI)~\cite{our}. At the full diamond on the horizontal
thin dotted line the linear increase of the thermal diffusivity has 
been experimentally found~\cite{WBM04} and at the full circle
$n_1^b\approx 1.3\times 10^{21}~{\rm cm}^{-3}$~\cite{our}
in good agreement with the experimentally estimated exciton
density $3.9\times 10^{21}~{\rm cm}^{-3}$~\cite{W01}. 
The open circles are the points where $-E_g(r_s,T)$ goes 
through the minimum. The energy gap of $\rm TmSe_{0.45}Te_{0.55}$ 
at ambient pressure is given by the vertical thin dotted line.
}
\label{f5}
\end{figure}

This can be seen in Fig.~\ref{f5} which combines 
the points where $-E_g(r_s,T)$ is 
minimal with (i) the phase boundary of the excitonic insulator, 
(ii) the $50\%$-contour of the bound state fraction 
$\gamma=n_1^b/(n_1^f+n_1^b)$ ($n_1^f$ and $n_1^b$ are the free and the bound 
part of the total density obtained from a T-matrix calculation of the total 
density), and (iii) the Mott line beyond which excitons disappear because Pauli 
blocking and screening lead to a vanishing exciton binding energy. All of it 
has been calculated in Ref.~\cite{our} within the static screening model. We 
expect however the quasi-static approximation to give similar results.

One of our main conclusions in Ref.~\cite{our} was that the resistivity 
anomaly, that is, the experimentally observed increase of the 
resistivity in a narrow pressure range, is most probably due to 
exciton-electron and exciton-hole scattering which would be very 
efficient above an excitonic insulator. Indeed, the temperature range, where the 
$50\%$-contour is located, coincides roughly with the temperature range
where the anomaly has been detected. In contrast, the analysis of the pair 
chemical potential performed in this paper suggests excitons in that region
to be unstable against a metallic electron-hole plasma. An increase 
of the resistivity is however hard to reconcile 
with a metallic phase. We expect therefore excitons in $\rm TmSe_{0.45}Te_{0.55}$ 
to be stabilised by band structure effects beyond 
the model shown in Fig.~\ref{f1} and additional scattering processes.
That the effective-mass, Wannier-type model may not be quite adequate can be also seen 
from the fact that  
the experimentally determined exciton Rydberg for $\rm TmSe_{0.45}Te_{0.55}$ 
is $\approx 75~{\rm meV}$ whereas the energy gap at ambient pressure is 
$\approx 135~{\rm meV}$
and thus only $\approx 1.8\cdot R_X$ (vertical thin dotted line in Fig.~\ref{f5}). 

In analogy to the hexaborides~\cite{ZR00}, we expect coherent inter-valley 
scattering of excitons to increase the exciton binding 
energy. Additionally, exciton-phonon scattering should also increase
the binding energy of an exciton in $\rm TmSe_{0.45}Te_{0.55}$ because 
the narrow VB makes the exciton very susceptible to phonon dressing. 
In fact, there is plenty of experimental evidence for strong 
exciton-phonon coupling in $\rm TmSe_{0.45}Te_{0.55}$~\cite{WBM04}.
Both scattering processes should be studied in detail, ideally
in a model which avoids the effective mass approximation and keeps the
full band structure. If it turns out that in a more realistic model 
the exciton binding energy is still 
above the chemical potential of the metallic phase, excitonic phases
must be ruled out in  $\rm TmSe_{0.45}Te_{0.55}$. On the other hand, if the 
exciton binding energy is pushed below the minimum of the metallic chemical potential, 
and that is not unrealistic considering the energy gain coherent inter-valley 
scattering alone can yield if the band structure permits it~\cite{ZR00}, excitons 
are stable, at least in some pressure range, and the excitonic instability leading 
to an excitonic insulator could take place in $\rm TmSe_{0.45}Te_{0.55}$ before 
an electron-hole gas-liquid transition interferes. 


\acknowledgment
Support from the SFB 652 is greatly acknowledged. We thank D. Ihle 
for valuable discussions.


\begin{thebibliography}{9}

\bibitem{HR68} For a review, see B. I. Halperin and T. M. Rice in: 
\textit{ Solid State Physics}, eds. F. Seitz, D. Turnbull, and H. 
Ehrenreich (Academic Press, New York, 1968) Vol. \textbf{21}, pp. 115. 

\bibitem{YHT99} D. P. Young, D. Hall, M. E. Torelli, Z. Fisk, J. L. Sarrao, J. D. 
Thompson, H.-R. Ott, S. B. Oseroff, R. G. Goodrich, and R. Zysler: 
Nature \textbf{397} (1999) 412.

\bibitem{ZRA99}
M. E. Zhitomirsky, T. M. Rice, and V. I. Anisimov: Nature \textbf{402} (1999) 251. 

\bibitem{BBM02}
E. Bascones, A. A. Burkov, and A. H. MacDonald: Phys. Rev. Lett. \textbf{89} (2002) 
086401.

\bibitem{WBM04} P. Wachter, B. Bucher, and J. Malar: Phys. Rev. B \textbf{69}
(2004) 094502.

\bibitem{NW90} J. Neuenschwander and P. Wachter: Phys. Rev. B \textbf{41} (1990) 12693; 
B. Bucher, P. Steiner, and P. Wachter: Phys. Rev. Lett. \textbf{67} 
(1991) 2717.  

\bibitem{W01} P. Wachter: Solid State Commun. {\bf 118} (2001) 645.

\bibitem{our} F. X. Bronold and H. Fehske, Phys. Rev. B \textbf{74} (2006) 165107.

\bibitem{R77} For a review, see T. M. Rice in: \textit{Solid State Physics}, 
eds. F. Seitz, D. Turnbull, and H. Ehrenreich (Academic Press, New York, 1977), 
Vol. \textbf{32}, pp. 1. 

\bibitem{HS84} H. Haug and S. Schmitt-Rink: Prog. Quant. Electr. \textbf{9} (1984) 3. 

\bibitem{ZR00} M. E. Zhitomirsky and T. M. Rice: Phys. Rev. B \textbf{62} (2000) 1492.

\bibitem{S77} R. N. Silver: Phys. Rev. B \textbf{8} (1973) 2403. 



\end{thebibliography}
\end{document}